\input phyzzx.tex
\tolerance=1000
\voffset=-0.0cm
\hoffset=0.7cm
\sequentialequations
\def\rl{\rightline}

\def\t1{{\tilde 1}}

\def\SUSY{supersymmetry }

\def\t{\theta}

\REF{\KIB}{M. B. Hindmarsh and T. W. B. Kibble, Rep. Prog. Phys. {\bf 58} (1995) 477, [arXiv:hep-ph/9411342].}
\REF{\POL}{E. J. Copeland, R. C. Myers and J. Polchinski, JHEP {\bf 0406} (2004) 013, [arXiv:hep-th/0312067]; J. Polchinski, [arXiv:hep-th/0412244].}
\REF{\BRA}{G. Dvali, Q. Shafi and S. Solganik, [arXiv:hep-th/0105203]; C. P. Burgess at. al. JHEP {\bf 0107} (2001) 047, [arXiv:hep-th/0105204].}
\REF{\EDI}{E. Halyo, JHEP {\bf 0403} (2004) 047, [arXiv:hep-th/0312268].}
\REF{\NIE}{H. B. Nielsen and P. Olesen, Nucl. Phys. {\bf B61} (1973) 45.}
\REF{\ABR}{A. A. Abrikosov, Sov. Phys JETP {\bf 5} (1957) 1174.}
\REF{\SUSY}{A. Achucarro and J. Urrestilla, JHEP {\bf 0408} (2008) 050, [arXiv:hep-th/0407193].}
\REF{\KAL}{G. Dvali, R. Kallosh and A. Van Proeyen, JHEP {\bf 0401} (2004) 035, [arXiv:hep-th/0312005].}
\REF{\KIBB}{T. W. B. Kibble, J. Phys. {\bf A9} (1976) 1387.}
\REF{\CMB}{L. Pogosian, S. H. H. Tye, I. Wasserman and M. Wyman, Phys. Rev. {\bf D68} (2003) 023506, [arXiv:hep-th/0304188]; L. Pogosian, I. Wasserman and M. Wyman, 
[arXiv:astro-ph/0403268].}
\REF{\PUL}{A. N. Lommen, [arXiv:astro-ph/0208572].}
\REF{\INS}{R. Blumenhagen, M. Cvetic, S. Kachru and T. Weigand, arXiv:0902.3251[hep-th] and references therein.}
\REF{\GEO}{C. Vafa, Journ. Math. Phys. {\bf 42} (2001) 2798, [arXiv:hep-th/0008142]; F. Cachazo, K. Intriligator and C. Vafa, Nucl. Phys. {\bf B603} (2001) 3, [arXiv:hep-th/0103067].}
\REF{\BCW}{R. Blumenhagen, M. Cvetic and T. Weigand, Nucl. Phys. {\bf B771} (2007) 113, [arXiv:hep-th/0609191].}
\REF{\IA}{L. Ibanez and A. Uranga, JHEP {\bf 0703} (2007) 052,[arXiv:hep-th/0609213]; JHEP {\bf 0802} (2008) 103, arXiv:0711.1316[hep-th].}
\REF{\IR}{L. Ibanez and R. Richter, JHEP {\bf 0903} (2009) 090, arXiv:0811.1583[hep-th].}
\REF{\CHR}{M. Cvetic, J. Halverson and R. Richter, arXiv:0905.3379[hep-th].}
\REF{\GW}{D. Green and T. Weigand, arXiv:0906.0595[hep-th].} 
\REF{\LAST}{E. Halyo, arXiv:0906.2159[hep-th].}
\REF{\QUI}{F. Cachazo, S. Katz and C. Vafa, [arXiv:hep-th/0108120].}
\REF{\DOUG}{M. Douglas, JHEP {\bf 9707} (1997) 004, [arXiv:hep-th/9612126].}
\REF{\SUP}{E. Witten, Nucl. Phys. {\bf B507} (1997) 658, [arXiv:hep-th/9706109]; M. Aganagic and C. Vafa, [arXiv:hep-th/0012041].}
\REF{\SHA}{M. Aganagic, C. Beem and S. Kachru, Nucl. Phys. {\bf B796} (2008) 1, arXiv:0709.4277[hep-th].}
\REF{\NPS}{S. Gukov, C. Vafa and E. Witten, Nucl. Phys. {\bf B584} (2000) 69, Erratum-ibid, {\bf B608} (2001) 477, [arXiv:hep-th/9906070].}
\REF{\TONG}{D. Tong, [arXiv:hep-th/0509216].}
\REF{\NAB}{M. Eto et. al., Phys. Rev. Lett. {\bf 96} (2006) 161601, [arXiv:hep-th/0511086]; Phys. Rev. {\bf D74} (2006) 065021; [arXiv:hep-th/0607070].}
\REF{\SEMI}{A. Achucarro and T. Vachaspati, Phys. Rev. {\bf D44} (1991) 3067; Phys. Rept. {\bf 327} (2000) 347, [arXiv:hep-ph/9904229].}
\REF{\AUZ}{R. Auzzi et. al., Nucl. Phys. {\bf B813} (2009) 484; arXiv:0810.5679[hep-th].}
\REF{\LIGO}{T. Damour and A. Vilenkin, Phys. Rev. Lett. {\bf 85} (2000) 3761, [arXiv:gr-qc/0004075]; Phys. Rev. {\bf D64} (2001) 064008, [arXiv:gr-qc/0104026].}
\REF{\DINF}{E. Halyo, Phys. Lett. {\bf B387} (1996) 43, [arXiv:hep-ph/9606423]; P. Binetruy and G. Dvali, Phys. Lett. {\bf B450} (1996) 241, [arXiv:hep-ph/9606342].} 
\REF{\SIN}{E. Halyo, JHEP {\bf 0407} (2004) 080, [arXiv:hep-th/0312042; hep-th/0402155]; [arXiv:hep-th/0405269].}
\REF{\TYE}{S. Sarangi and S. H. Tye, Phys. Lett. {\bf B536} (2002) 185, [arXiv:hep-th/204074]; N. T. Jones, H. Stoica and S. H. Tye, Phys. Lett. {\bf B563} (2003) 6, 
[arXiv:hep-th/0303269]; G. Dvali and A. Vilenkin, [arXiv:hep-th/0312007]; K. Dasgupta, J. Hsu, R. Kallosh, A. Linde and M. Zagerman, JHEP {\bf 0408} (2004) 030, [arXiv:hep-th/0405247]; 
T. Matsuda, Phys. Rev. {\bf D70} (2004) 023502, [arXiv:hep-ph/0403092].}
\REF{\MAL}{S. Kachru et. al.  JCAP {\bf 0310} (2003) 013, [arXiv:hep-th/0308055].}
\REF{\HAN}{A. Hanany and D. Tong, JHEP {\bf 0404} (2004) 066, [arXiv:hep-th/0403158].}
\REF{\MOD}{M. Eto et. al., Phys. Rev. {\bf D76} (2007) 105002, arXiv:0704.2218[hep-th].}
\REF{\INT}{M. G. Jackson, N. T. Jones and J. Polchinski, JHEP {\bf 0510} (2005) 013, [arXiv:hep-th/0405229].}
\REF{\REC}{M. Eto et. al., Phys. Rev. Lett. {\bf 98} (2007) 091602, [arXiv:hep-th/0609214].}

\singlespace
\rl{SU-ITP-09/29}
\pagenumber=0
\normalspace
\medskip
\bigskip
\titlestyle{\bf{Cosmic Strings with Small Tension}}
\smallskip
\author{ Edi Halyo{\footnote*{e--mail address: halyo@stanford.edu}}}
\smallskip
\centerline {Department of Physics} 
\centerline{Stanford University} 
\centerline {Stanford, CA 94305}
\smallskip
\vskip 2 cm
\titlestyle{\bf ABSTRACT}
We describe cosmic F--term strings with exponentially small tension which are D3 branes wrapped on deformed $A_3$ singularities. We show that brane instanton 
effects which can be calculated after a geometric transition give rise to an exponentially small volume for the node on which the D3 branes wrap leading to a string with small tension. 
We generalize our description to the case of non--Abelian cosmic strings and argue that these strings are stable against monopole--anti monopole pair creation.
\singlespace
\vskip 0.5cm
\endpage
\normalspace

\centerline{\bf 1. Introduction}
\medskip

The renewed interest in cosmic strings[\KIB] is due to the fact that their detection may provide a striking signature of string theory 
which contains many objects that can play the role of cosmic strings. The simplest possibility is that cosmic strings are either fundamental strings or D1 branes, i.e. F or D strings[\POL].
These are especially relevant in the context of D--brane inflation[\BRA] since they are produced at the end of brane--anti brane annihilation. Alternatively, cosmic strings may be D(p+1)
branes wrapped on p--cycles of the compact space[\EDI].

In field theory, cosmic strings (or vortices) exist in models with a spontaneously broken $U(1)$ gauge symmetry[\NIE,\ABR].
In supersymmetric field theories, vortex solutions arise from nonzero F or D--terms and are called F or D--term strings respectively[\SUSY]. In ${\cal N}=2$ supersymmetric models 
both types of strings are BPS; however with ${\cal N}=1$ supersymmetry, only D--term strings are BPS. F--term strings are topologically stable but
their solutions and tensions receive corrections[\KAL].

If cosmic strings were produced before inflation, they would be diluted and not be observable today. Therefore, we must assume that
they were produced after inflation. In a cosmological context, cosmic strings are produced by the Kibble mechanism[\KIBB] which arises from the finite horizon size. 
Cosmic strings can be detected mainly
due to their gravitational effects such as gravitational lensing, contributions to the CMB density perturbations and gravitational radiation from string loops[\KIB]. The present bounds 
on their tension, $T_s$, arise either from the density fluctuations in the CMB[\CMB] with $GT_s<2 \times 10^{-7}$ or pulsar timing[\PUL] with
$GT_s<1.3 \times 10^{-7}$. Thus, the string tension must be suppressed with respect to the string scale, $M_s \sim M_P$. This is not a problem in field theory since the 
scale of spontaneous $U(1)$ breaking is arbitrary (at the cost of
introducing a gauge group at intermediate scales). In string theory, on the other hand, the smallness of the tension is difficult to explain. One possibility is to locate cosmic strings
in a warped throat which effectively lowers their observed tension[\POL].

In this letter, we consider the $3+1$ dimensional world--volume theory of D5 branes on a deformed $A_3$ singularity (with three nodes).
We show that this world--volume theory contains a vortex which is an F--term string. We give a different description of the same cosmic string as a D3 brane wrapped 
on a node of the singularity. The string tension is exponentially small due to brane instanton effects[\INS] on another node of the singularity which can be calculated after a 
geometric transition[\GEO]. (For other phenomenological consequences of brane instantons see [\BCW-\LAST].)
In the world--volume theory, this instanton induces an exponentially small F--term which results in an exponentially small string tension. 
Alternatively, the brane instanton results in an exponentially small ``stringy volume" of the node which the D3 brane wraps giving rise to a small string tension.

A simple generalization of the model, in which there are multiple D5 branes wrapped on each node of the deformed $A_3$ singularity leads to non--Abelian cosmic strings with small 
tension. We also show that these cosmic strings are stable against monopole--anti monopole pair creation.  

\bigskip
\centerline{\bf 2. Cosmic Strings with Small Tension}
\medskip

In this section, we describe the physics on the world--volume of D5 branes wrapped on a deformed $A_3$ singularity. In section 2.1 we describe cosmic strings as F--term strings in 
field theory or as D3 branes wrapped on a node of the singularity. In section 2.2 we generalize the results of the previous section to non--Abelian cosmic strings. In 
section 2.3 we consider the stability of cosmic strings with respect to monopole--anti monopole pair creation.

{\bf 2.1. F--term Strings with Small Tension}: 
Consider the deformed $A_3$ singularity given by
$$uv=(z-mx)(z+mx)(z+mx)(z-m(x-2a)) \eqno(1)$$ 
with three nodes (singular $S^2$s). The $A_3$ singularity is fibered over the complex plane $C(x)$. 
We see that the first node is at $x=0$ whereas the third one is at $x=a$. The location of the second node is not fixed since it corresponds to a flat direction on the world--volume theory. 
For the moment, we take its location to be at $x \not = 0,a$ so that the three nodes are separate.
We wrap one D5 brane on each node and consider the theory living on the noncompact $3+1$ dimensional D5 brane world--volume.
(We could wrap $N_1$ and $N_3$ D5 branes on the first and third nodes with minor changes in our results. The important point is to have one D5 brane wrapped on the second node in
order to have a $U(1)$ gauge group that gives rise to the vortex.)
The world--volume gauge group is $U(1)_1 \times U(1)_2 \times U(1)_3$ with the couplings[\QUI]
$${{4 \pi} \over g_{i}^2}={V_i \over {(2 \pi)^2 g_s \ell_s^2}} \qquad i=1,2,3 \eqno(2)$$
where $g_s$ and $\ell_s$ are the string coupling and length respectively and $V_i$ is the ``stringy volume"[\DOUG] of the $i$th node given by $V_i=(2 \pi)^4 \ell_s^4(B_i^2+r_i^2+\alpha_i^2)^{1/2}$. Here
$$B_i=\int_{S^2} B^{NS}   \qquad  r_i^2=\int_{S^2} J \eqno(3)$$
i.e. $B_i$ is the NS-NS flux through the $i$th node and $r_i^2$ is the volume of the blown-up $S_i^2$s. On the second node, we set both of these to zero. 
Thus, for the moment there is no flux
or small resolution on the second node. However, there are (complex) deformations of the singularity parametrized by $\alpha_i$ which result in nonzero ``stringy volumes'' for all the nodes. 
These are related to the deformations of the superpotential, i.e. singlet masses and F--terms obtained from[\SUP]
$$W(\phi_i)= \int^{\phi_i} (z_i(x)-z_{i+1}(x)) dx \eqno(4)$$
where $z_i(x)$ are the zeros of the different factors in eq. (1) which describe the deformation of the singularity. In general, we may have fluxes on the fist and third nodes so that
$V_i$ are all different. 

The matter content on the world--volume consists of three singlets $\phi_{1,2,3}$ and two pairs of bifundamentals $Q_{12},Q_{21},Q_{23},Q_{32}$ coupled through the superpotential[\SUP]
$$W=(\phi_2-\phi_1)Q_{12}Q_{21}+(\phi_3-\phi_2)Q_{23}Q_{32}+m\phi_1^2-m(\phi_3-a)^2 \eqno(5)$$
This assumes a certain normalization of the $\phi_i$ on which we will elaborate later. We see that a supersymmetric vacuum exists for vanishing bifundamental VEVs where the singlet VEVs 
are $\phi_1=0$, $\phi_3=a$ and $\phi_2$ is free. This exactly matches the locations of the nodes mentioned above.

We would like to consider the brane instanton correction[\INS] to the above theory where the instanton in question is a Euclidean D1 brane wrapped on the third node[\SHA]. 
This effect can be reliably
computed after a geometric transition[\GEO] if the third node is isolated and the fields living there are all massive (which hold in our case). The nonperturbative instanton effect before the
transition is given by the perturbative flux superpotential after the transition.
Consider a geometric transition at the third node with $S^2_3 \to S^3$ where the D5 brane wrapped on the third node is replaced by a unit RR flux
$$N=\int_{S^3} H^{RR}=1 \eqno(6)$$
The geometry becomes[\SHA]
$$uv=(z-mx)(z+mx)((z+mx)(z-m(x-2a))-s) \eqno(7)$$ 
where $s=mS$ and $S$ is the size of the blown up $S^3$. After the geometric transition, the fields that live on the third node $\phi_3, Q_{23}, Q_{32}$ (and $U(1)_3$) disappear and 
the superpotential gets two new contributions. The first one is the flux superpotential (for $N=1$)[\NPS]
$$W_{flux}={V_3 \over {2 \pi g_s \ell_s^2}} S+ \partial_S {\cal F}_0 \eqno(8)$$
where ${\cal F}_0$ is the prepotential. 
The form of the superpotential in this case is known to be (in notation more common in the literature $t=V/2 \pi \ell_s^2$)[\GEO]
$$W_{flux}={V_3 \over {2 \pi g_s \ell_s^2}}S+ S \left(log {S \over \Delta^3} -1 \right) \eqno(9)$$
The second instanton contribution is the correction to the superpotential of $\phi_2$ given by[\SHA]
$$W^{\prime}(\phi_2)=\int^{\phi_2} (z_2-{\bar z}_3) dx \eqno(10)$$
where ${\bar z}_3$ is the solution to
$$(z+mx)(z-m(x-2a))=s \eqno(11)$$
given by ${\bar z}_3=-ma- \sqrt{m^2(x-a)^2+s}$. Then we find
$$W^{\prime}(\phi_2)=-{S \over 2} log {|\phi_2-a| \over \Delta} \eqno(12)$$
For small $\phi_2$ (which we will justify later) this is equivalent to an F--term
$$F_{\phi_2}={\partial W^{\prime} \over \partial \phi_2}=-{S \over {2a}} \eqno(13)$$
Note that $F<0$. The superpotential now becomes
$$W=m\phi_1^2+(\phi_2-\phi_1)Q_{12}Q_{21}+\phi_2 F+W_{flux} \eqno(14)$$
where $W_{flux}$ and $F$ are given by eqs. (9) and (13). At low energies, $E<<m$, $\phi_1$ decouples. Its F--term is
$$F_{\phi_1}=2m\phi_1-Q_{12}Q_{21} \eqno(15)$$
Setting this to zero gives $\phi_1=Q_{12}Q_{21}/2m$ so $\phi_1$ decouples with a VEV that depends on (the VEVs of) $Q_{12},Q_{2}$. The F--term above also induces a nonrenormalizable term 
in the superpotential of the type $(Q_{12}Q_{21})^2/4m$. 

The field $S$ also decouples at low energies. Setting $F_S=0$ we find
$$S=S_0=\Delta^3 e^{-V_3/{2 \pi g_s \ell_s^2}} \eqno(16)$$
which is exponentially small because it is due to an instanton effect. From eq. (13) we find that
$$F_{\phi_2}=-{S_0 \over {2a}}=-{\Delta^3 \over {2a}} e^{-V_3/{2 \pi g_s \ell_s^2}} \eqno(17)$$
thus obtaining an exponentially small F--term. We can also decouple the $U(1)_1$ gauge field from matter by taking $g_1^2$ to be very small. From eq. (2) we see that this can be 
done by taking $V_1>>\ell_s^2$. Thus, $U(1)_1$ becomes a global symmetry and the only gauge group left is $U(1)_2$.
The remaining low--energy superpotential is
$$W=\phi_2 Q_{12}Q_{21}-{{(Q_{12}Q_{21})^2} \over {4 m}}+\phi_2 F \eqno(18)$$
This leads to the scalar potential
$$V_F= |Q_{12}Q_{21}+F|^2+|\phi_2|^2(|Q_{12}|^2+|Q_{21}|^2)+ O(Q^4/m^2) \eqno(19)$$
We note that the terms of $O(Q^4/m^2)$ which we did not write out explicitly are important for obtaining the correct VEV of $\phi_2$.
Under $U(1)_2$, the fields $Q_{12},Q_{21}$ have charges $+1,-1$ respectively. As a result, there is a D--term contribution to the scalar potential
$$V_D={g_2^2 \over 2} (|Q_{12}|^2-|Q_{21}|^2)^2 \eqno(20)$$
The total scalar potential is given by $V=V_F+V_D$, i.e the sum of eqs. (19) and (20). $V_D$ is minimized by $|Q_{12}|=|Q_{21}|$ so we can take $Q_{12}=Q_{21}^{\dagger}$, i.e.
the two charged fields have complex conjugate VEVs. In addition, from eqs. (15) and (18) (taking into account the terms of order $O(Q^4/m^2)$) we find that the F--terms vanish 
for $\phi_1=\phi_2=-F/2m$. From eq. (17) we see that $\phi_2<<a$ justifying our assumption above.

For $E<\sqrt{|F|}$, $\phi_2$ decouples and we are left with a theory described by the
Lagrangian 
$$L=-{1 \over {4 g_2^2}} F_{\mu \nu} F^{\mu \nu}-{1 \over g_2^2} (D_{\mu}Q_{12}D^{\mu}Q_{12}^{\dagger}-D_{\mu}Q_{21}D^{\mu}Q_{21}^{\dagger})- |Q_{12}Q_{21}+F|^2 \eqno(21)$$
where $D_{\mu}Q=(\partial_{\mu}-igA_{\mu})Q$ and $F_{\mu \nu}=\partial_{\mu}A_{\nu}-\partial_{\nu}A_{\mu}$. This theory
has a vortex solution; in fact this is known as an F--term string[\SUSY]. 

Far from the core of the vortex, at large $r$, the solution is 
$$ Q_{12}=Q_{21}^{\dagger}=\sqrt{|F|}e^{i n \theta}  \qquad  A_{\theta}={n \over {gr}} \qquad  F_{\mu \nu}=0 \eqno(22)$$
where $n$ is the topological winding number. The vortex (along the $z$ direction) has a metric which has a conical singularity ($F<0$)
$$ds^2=-dt^2+dz^2+dr^2+r^2 \left (1+{{nF} \over M_P^2} \right)d \theta^2 \eqno(23)$$
Near the core of the vortex, at small $r$, the solution is
$$Q_{12}=Q_{21}=0 \qquad  A_{\theta}={M_P^2 \over {gF}} \left(1-cos \left({{gF} \over M_P} \right)r \right) \qquad  F_{r \theta}=-M_P sin \left({{gF} \over M_P} r \right) \eqno(24)$$
It is well known that the vortex with winding number $n$ carries a magnetic flux of
$$\Phi_n=\int B_z dxdy=2 \pi n \eqno(25)$$
which shows that the topological charge is magnetic flux. The string has tension
$$T_n=2 \pi |F| n={{\pi \Delta^3 } \over a} n e^{-V_3/{2 \pi g_s \ell_s^2}} \eqno(26)$$

We see that the above tension is exponentially smaller than the string scale $M_s$ (or $M_P$). For example, if we take $\Delta \sim a \sim M_s$ then the bound on the tension 
becomes $G T_S \sim (\pi/g_s^2) e^{-V_3/2 \pi g_s \ell_s^2}<10^{-7}$. For $g_s \sim 1/2$, this means (with $r_3=\alpha_3=0$) 
$V_3 > 60 \ell_s^2$ which following eq. (3) requires $\int_{S_3^2} B^{NS} > (1/10) \ell_s^2$. Note that the string instanton factor is precisely what one would expect from a field theory 
instanton (for $U(1)_3$) 
$$exp(-V_3/2 \pi g_s \ell_s^2)=exp(-8 \pi^2/g_{YM}^2) \eqno(27)$$ 
where we used eq. (2). It is difficult to calculate this instanton effect in the world--volume field theory. However, the instanton corrections to the superpotential given by eqs. (9) and (12) 
can be easily calculated in string theory using a geometric transition.

The F--term string that we found in the world--volume field theory is in fact a D3 brane wrapped on $S_2^2$[\EDI]. We can find the
relation between the F--term in eq. (17) and the ``stringy volume'' of the second node $V_2$ by equating the energy of the D5 brane wrapped on $S_2^2$ to the vacuum energy in the field theory,
$${1 \over 2} g_2^2 F^2= T_{D5} V_2={V_2 \over {(2 \pi)^5 g_s \ell_s^6}} \eqno(28)$$
Note the extra factor of $g_2^2/2$ above. This is due to the normalization of $\phi_2$ that is different fom that in eq. (18) (which is the common one in the literature). In fact, there is a 
normalization in which all $\phi_i$ are multiplied by $g_i/ \sqrt {2}$ in the superpotential given by eq. (5) which is what is obtained from the world--volume theory. In order to find 
the relation
between $F$ and $V_2$ it is crucial to use this normalization. Since we took $g_1$ to be very small above (to decouple $U(1)_1$) we need to make sure that $g_1^2 m$ and $g_1^2 a$ are finite
which can be easily done. Since this subtlety does not affect the rest of our results we keep the common normalization of $\phi_i$ everywhere else.

Using eq. (2) for $g_2$ we find
$$|F|={V_2 \over {(2 \pi)^4 g_s \ell_s^4}} \eqno(29)$$
which establishes the relation between $F$ and $V_2$. Since $F$ is exponentially small the same is true for $V_2$ in string units
$$V_2=(2 \pi)^4 g_s \ell_s^4 \left(\Delta^3 \over {2a} \right) e^{-V_3/2 \pi g_s \ell_s^2} \eqno(30)$$
We can now find the tension of a D3 brane wrapped on $S_2^2$
$$T_s=T_{D3} V_2={V_2 \over {(2 \pi)^3 g_s \ell_s^4}}=2 \pi |F| \eqno(31)$$
This shows that the F--term string we found in field theory is a D3 brane wrapped on $S_2^2$. The smallness of the string tension is a direct result of the exponentially small volume
$V_2$. It is well--known that a D3 brane inside a D5 brane constitutes a generalization of a magnetic flux tube due to the coupling between the RR potential and the world--volume gauge 
field strength. After wrapping both branes on $S_2^2$, the wrapped D3 brane carries one unit
of magnetic flux as expected from a cosmic string. The topological charge $n$ is simply the number of times the D3 brane wraps $S_2^2$.

From eqs. (2) and (30) for $g_2^2$ and $V_2$ we find that the $U(1)_2$ coupling is exponentially large
$${{4 \pi} \over {g_2^2}}=(2 \pi)^2 \ell_s^2 \left(\Delta^3 \over {2a} \right) e^{-V_3/2 \pi g_s \ell_s^2} \eqno(32)$$
Therefore, the wrapped D3 brane corresponds to the vortex solution in the very strong coupling limit. This is a direct result of the fact that the only contribution to the volume of $S_2^2$
comes from the brane instanton effect. As we will discuss in section 2.3, if the second node already has a large volume, e.g. due to a nonzero flux $B_2$, we can get $g_2^2<<1$.

In general, in this limit vortices are thin, i.e. they are much thinner than
the characteristic scale defined by the F--term (which in our case is exponentially smaller than $\ell_s$). In our case, the vortex width is
$$w \sim {1 \over {g_2 \sqrt{|F|}}} \sim \pi \ell_s \eqno(33)$$
which is independent of the field theory parameters. This is a direct result of the identical dependence of $g_2^{-2}$ and $F$ on the exponentially
small volume $V_2$.


It is interesting to note that the low--energy superpotential in eq. (18) has ${\cal N}=2$ supersymmetry in the limit $m \to \infty$ or for very low energies even though we started 
from a geometry and superpotential that had only ${\cal N}=1$ supersymmetry. However, for large but finite $m$, this superpotential has only ${\cal N}=1$ supersymmetry due to the 
nonrenormalizable interactions arising from integrating $\phi_1$ out. In addition, eq. (13) takes into account only the lowest term in the expansion
of the stringy correction in eq. (12). Clearly, there are other exponentially suppressed terms which are higher order in $\phi_2$. We neglected these corrections to eq. (18)
since they are small. Thus we find that in our model, ${\cal N}=2$ supersymmetry is broken explicitly down to ${\cal N}=1$ by nonrenormalizable and exponentially suppressed terms.
In models with ${\cal N}=2$ supersymmetry, F--term strings are BPS and therefore their solution is exact[\SUSY]. On the other hand, with only ${\cal N}=1$ supersymmetry, F--term strings are not BPS and therefore 
their solution and tension receive corrections[\KAL]. We expect these corrections to be inversely proportional to the large mass $m$ and/or exponentially suppressed.
Cosmic strings that correspond to these slightly modified solutions are expected to be stable due to conservation of topological charge. Therefore, the solution in eqs. (22)-(24)
is stable even though it gets corrected.

\medskip
{\bf 2.2. Non--Abelian F--term Strings}:
We can also obtain non--Abelian vortices[\TONG,\NAB] by wrapping multiple D5 branes on the nodes of the $A_3$ singularity. If we wrap $N_3$ D5 branes on the third node, after the geometric transition
we have $N_3$ units of RR flux as in eq. (6). Then, the flux superpotential becomes
$$W_{flux}={V_3 \over {2 \pi N_3 g_s \ell_s^2}}S+ S \left(log {S \over \Delta^3} -1 \right) \eqno(34)$$
where the only change is the factor of $N_3$ in the denominator. As a result, the VEV $S_0$ becomes
$$S_0=\Delta^3 e^{-V_3/{2 \pi N_3 g_s \ell_s^2}} \eqno(35)$$
Of course, this change in the exponent can be compensated by taking the volume $V_3$ to be $N_3$ times larger than that in the previous section. Therefore, the only effect of wrapping $N_3>1$
branes on the third node is to require a larger volume, $V_3$, for the same exponential suppression of the tension. 

In order to get a non--Abelian vortex, we can wrap $N_f$ and $N_c$ D5 branes on the first and second nodes respectively giving rise to a $U(N_f)\times U(N_c)$ gauge group.
Then, the bifundamentals $Q_{12}$, and $Q_{21}$ are in the $(N_f, {\bar N_c})$ and $({\bar N_f}, N_c)$ representations. 
If we take the volume of the first node, $V_1$, to be very large, i.e. $V_1>>\ell_s^2$ the $U(N_f)$ coupling given by
$${{4 \pi} \over g_1^2}={V_1 \over {(2 \pi)^2 g_s \ell_s^2}}  \eqno(36)$$
becomes very small. Therefore, the gauge dynamics decouples and $U(N_f)$ becomes a global symmetry. $Q_{12},Q_{21}$ become $N_f$ flavors in the $N_c$ and ${\bar N_c}$ representations
of the gauge group $U(N_c)$. The singlet $\phi_2$ is now an adjoint of the local $U(N_c)$ and a singlet of the global group $U(N_c)$.
 
There is no string solution for $N_f<N_c$ since the F--constraints cannot be satisfied due to the rank condition. On the other hand, for $N_f>N_c$ strings become  
semi--local[\TONG,\SEMI] (in the language of $N_c=1$) and may have arbitrarily large size[\AUZ]. Therefore, stability of strings requires $N_f=N_c=N$.  We can realize this by wrapping
an equal number of D5 branes on the first two nodes. We end up with $N$ flavors of $Q_{12},Q_{21}$ in the $N, {\bar N}$ of the gauged $U(N)$ and a gauge singlet $\phi_2$ in the adjoint
of the flavor group. 
The physics is described by the generalization of eq. (21) for $U(N)$ with $N_f=N$ flavors. The solutions are trivial generalizations of eqs. (22)--(24) which 
describe non--Abelian strings[\TONG].
The non--Abelian F--string tension is the same as the Abelian one $T_s=2 \pi |F|$ where $F$ is
$$F=-{\Delta^3 \over {2a}} e^{-V_3/{2 \pi N_3 g_s \ell_s^2}} \eqno(37)$$
giving an exponentially small tension as before. In order to satisfy the constraint $GT_s<10^{-7}$ we now need a bigger volume, $V_3>60 N \ell_s^2$.
It is clear that non--Abelian cosmic strings are also a D3 branes wrapped on the second node.

\medskip
{\bf 2.3. Stability of Strings}:
Cosmic strings may be unstable due to monopole--anti monopole creation. If the rate for pair creation is not too small, then monopole--anti monopole pairs are created on the string which
breaks into small pieces. In fact, in our model there are monopoles and we have to make sure they do not destabilize the cosmic string. Outside the string, the supersymmetric 
vacuum is given by eq. (22) with $\phi_1=\phi_2=-F/2m$. Near the core of the string, on the other hand, $Q_{12}=Q_{21}=\phi_1=0$ and $\phi_2$ is free. Therefore, for $\phi_2 \not =0$
there are monopoles that live inside cosmic strings, i.e. the monopoles are confined by the strings which are magnetic flux tubes. These monopoles have mass
$$m_m={{4 \pi} \over g_2^2} |\phi_2| \eqno(38)$$
On the other hand, these monopoles are simply D3 branes wrapped on the second node with volume $V_2$ in eq. (30) and stretched between the D5 branes wrapped on the first and second nodes. 
They have mass
$$m_m=T_{D3} V_2 d_{12}={V_2 \over {(2 \pi)^4 g_s \ell_s^4}} \phi_2 (2 \pi \ell_s^2) \eqno(39)$$
which matches the monopole mass in eq. (38). The monopole--anti monopole pair creation probability is 
$$P \sim e^{- \pi m_m^2/T} \sim exp[-(8 \pi^2 \phi_2^2) / (g_2^4 |F|)] \eqno(40)$$
Since $g_2^4 F$ is exponentially large (see eqs. (29), (30) and (32)) we find $P=1$ which means that the string is not stable. It breaks very fast
into small pieces due to monopole-anti monopole pair creation. This is a direct result of the fact that $V_2$ and therefore $g_2^{-2}$ is exponetially small
due to the brane instanton effect. Previously, for simplicity, we assumed that there was no flux or blow--up on the second node. If we change this assumption,
e.g. assume a nonzero flux $B_2$, we can easily obtain a small coupling
$${{4 \pi} \over {g_2^2}}={{(2 \pi)^2 \ell_s^2} \over g_s} B_2 \eqno(41)$$
for large enough $B_2$ given by eq. (3). In this case, the brane instanton effect changes the nonzero $V_2$ by an exponentially small amount, $\Delta V_2$.
This, in turn, changes the energy of the D3 brane wrapped on $S_2^2$ by an exponentially small amount which, as before, we identify with the F--term contribution to the energy in the field 
theory. 
Thus, $F$ is still exponentially small whereas $g_2^2$ is $O(1)$. Using eqs. (41) and (42) we find that $P=0$ to great accuracy. This shows that
the cosmic strings we described above can be made stable quite easily by assuming a nonzero volume for the second node before noperturbative
effects are taken into account.

\bigskip
\centerline{\bf 3. Conclusions and Discussion}
\medskip

We see that a D3 brane wrapped on a node of a deformed $A_3$ singularity is a good cosmic string candidate. In field theory, this corresponds to an F--term string. The string has an exponentially 
small tension due to brane instanton effects on a node of the singularity. These nonperturbative effects are calculable after a geometric transition on that node. 
They give rise to an exponentially small volume on which the D3 brane wraps resulting in a small tension.
In field theory, this manifests itself as an exponentially small F--term.
By wrapping multiple D5 branes on the nodes of the singularity, we obtain non--Abelian strings with exponentially small tension.

If cosmic strings are D3 branes wrapped on singularities, $A_3$ which has three nodes is the smallest singularity that leads to strings with a small tension. One node is needed to wrap
the D3 brane on (or to have the Abelian gauge group). Another node is required in order to have the scalars $Q_{12},Q_{21}$ that break the $U(1)$ spontaneously since there is a pair of 
bifundamentals for each link between two neighboring nodes. Finally, we need a third node on which there is a brane instanton and which goes through a geometric transition.

It is expected that LIGO and LISA will be able to detect cosmic strings with $GT_s \sim 10^{-(10-12)}$ in the near future[\LIGO]. We showed above that
our model can easily accomodate cosmic strings with such small tensions. This possibility for detection leaves us with a few important questions.
The first one is whether it is possible to build D--brane inflation models on singularities that end in a configuration similar to the one we started with. These would probably need
to be D--term inflation models[\DINF] that avoid the well--known inflaton mass problem. In fact, models of inflation on singularities similar to the one we used in this paper have been
constructed[\SIN]. However, it is not clear whether cosmic strings are produced after inflation since in these models since there is no brane--anti brane annihilation. Nevertheless,
cosmic strings may be produced if the (factional) branes get close enough during inflation.

On the other hand, it is known that cosmic strings[\TYE] can be produced after D--brane inflation due to brane--anti brane annihilation[\BRA,\MAL]. However, there is 
no reason for these to have small tensions except when they are produced in warped spaces[\POL]. In that case, the string tension gets redshifted by the warp factor and can be
exponentially small. Brane instanton effects on singular spaces provide an alternative to the warped space scenario for producing strings with exponentially small tension. 

Second, we need to know the details of the $1+1$ dimensional string world--sheet theory in order to understand string loop dynamics. This is cruial for calculating cusp effects 
that are the source of gravitational radiation from string loops which are detected in pulsar timing measurements[\KIB]. The string world--sheet theory has been obtained in the context of 
intersecting brane models[\HAN,\MOD]. It is important to find out if the same or a similar string world--sheet theory describes D3 branes wrapped on singularities.

Finally, if a cosmic string is detected, what would be the signature that distinguishes between a field theory vortex and a wrapped D3 brane? One possibility is the
interaction between the cosmic strings, i.e. the probability for intercommutation which has different properties for different types of strings[\INT]. For example, even though the cosmic strings
described above are wrapped D3 branes, we expect the probability for intercommutation when two strings cross each other to be $P \sim 1$ since they are also field theory vortices[\REC].
Thus, in this respect we do not expect a difference. However, the
wrapped D3 branes behave like field theory vortices only at low energies $E<<m$ at which we can neglect the massive fields and 
nonrenormalizable interactions in the field theory. At higher energies, the brane world--volume theory deviates from a gauge theory and is described by a Born--Infeld theory. 
It would be interesting to investigate these differences and their observational signatures.

\bigskip

\centerline{\bf Acknowledgements}

I would like to thank the Stanford Institute for Theoretical Physics for hospitality.

\vfill

\refout

\end
\bye